# The Pilot Alignment Pattern Design in OFDM Systems

*,1Yong Chan Lee, 2Won Chol Jang, 3Un Kyong Choe, Gyong Chol Leem

1,3College of Computer Science, Kim Il Sung University, D.P.R.K
Email: 1leeyongchan@yahoo.com
Email: 3 ChoeUnKyong@yahoo.com
2,4Information Center, Kim Il Sung University, D.P.R.K
Email: 2univjwc@yahoo.com
Email: 4leemilun@126.com

*Abstract*— In this paper, we propose optimal pilot pattern of downlink OFDM (Orthogonal Frequency Division Multiplexing) communication system. Current pilot patterns in OFDM system are evaluated with new criteria, new pilot pattern and "distance" filter which has small computational burden is suggested. The combination of suggested cell pilot pattern and "distance" filter provides better performance than LMMSE estimation which implements time-frequency correlation about some patterns.

Keywords- pilot pattern, OFDM

## I. INTRODUCTION

Orthogonal frequency division multiplexing (OFDM) is a promising technique for achieving high data rate and combating multi-path fading in wireless communications. [1] For further improvement of the transmission performance, the channel impulse response (CIR) estimation is required, especially for coherent detection. Pilot symbol aided channel estimation (PSAE) is commonly used in OFDM systems [2-4].

The CIR can be estimated with predetermined pilot symbols in real time. In OFDM system, the pilot symbols are scattered in the time and frequency domain to track time-variant and frequency-selective channel characteristics. The CIR for coherent detection of data symbol can be obtained by interpolating the CIR estimation with pilot symbols. The optimum interpolation can be performed using a two dimensional (2-D) Wiener interpolator with infinite taps. The mean square error (MSE) of the Wiener channel estimation depends not on the pattern but only on the density of pilot pattern [5]. However, it may not be practical to use such a Wiener interpolator because of large implementation complexity. As a result, simple interpolators, such as linear, Lagrange or Spline interpolator, are often employed in practice [6, 7]. When these interpolators are employed, the performance of channel estimation is affected by the pilot pattern (i.e., the shape and spacing) as well as the pilot density [8-11]. For example, when the channel is fast time-variant with a small multi-path delay spread, it would be advantageous to insert more pilot symbols in the time domain than in the frequency domain. Most of previous researches on the design of pilot symbol for channel estimation in mobile OFDM systems have been obtained based on the computer simulation results [8-11]. Given a pilot density, the shape and spacing of the optimum pilot pattern has been derived by minimizing the MSE of the estimated CIR with the use of a conventional interpolator in [12].

The alignment of pilot symbol and the selection of pilot sequence affect the performance of channel estimation largely. In general, the alignment of pilot symbol is specified by time delay and Doppler frequency, that is, characteristic parameters of channel. Likewise, the selection of pilot sequence is the main factor in eliminating interference between pilot sequence and inter-path interference canceling (IPIC). Which antenna of multiple transmission antennas will be used to transmit which training sequence under the background of MIMO-OFDM system was investigated in literature [16].

In this paper, we propose a new criterion of pilot pattern evaluation, derive optimal pilot pattern based on it and propose pilot pattern adaptive method. First, we propose Cell pilot pattern-improved form of Diamond pilot pattern. Then we propose distance filter suitable for Cell pilot pattern.

The paper is organized as follows. In section 2 various pilot patterns are considered, criteria for pilot pattern are derived and optimal pilot pattern is proposed. The effectiveness of the proposed methods is considered in section 3 and section 4 is conclusion.

## II. SUGGESTED METHODS

### A. *Various pilot patterns*

For a given pilot density, the spacing of optimal pilot pattern that minimizes MSE of estimated CIR with the use of a conventional interpolator that was derived in [12] is as following.

$$\overline{w_1}^{(4)} d_t^4 = \overline{w_2}^{(4)} d_f^4 \qquad (1)$$

where $\overline{w_1}^{(4)}$ is the fourth-order moment of the Doppler Spectrum, $\overline{w_2}^{(4)}$ is the fourth-order moment of the power delay profile, $d_f$ is the pilot spacing in the time and is the pilot spacing in the frequency. Besides, pilot spacing in the time and frequency should satisfy following conditions. [13]

$$f_{\max} T_{sym} d_t \leq \frac{1}{2}, \quad d_f N_T \tau_{\max} / T \leq 1 \qquad (2)$$

where $f_{\max}$ is the maximum Doppler frequency, $T_{sym} = (N_{FFT} + N_{cp})T_{spl}$ is the symbol duration with guard interval $T = N_{FFT} T_{spl}$ is the symbol duration without guard interval, $N_T$ is the number of transmitting antennas, $T_{spl}$ is the sample duration, $f_d$ is the bandwidth of Doppler frequency, and $\tau_{\max}$ is the maximum delay of the channel.

The typical pilot patterns in the $D-2$ time-frequency grid for OFDM systems are as following: [12, 13, and 14]

The choice of 4) in literature [1], 5) in [2] and 6) in [3] were optimal, respectively. It is also important to determine what type of estimation and interpolation are to be used for a given pilot density. In LMMSE method, which requires channel covariance matrix, estimation of channel covariance matrix based on second-order statistics should be done. Therefore, in practical communication systems, it gives high calculation burden. So lots of systems use LS method to interpolate with estimation on the pilot symbol [13].

There are several methods for first-order interpolation [15]: LI (Linear Interpolation), SOI (Second-Order interpolation), LPI (Low-Pass filter Interpolation), SCI (Spline Cubic Interpolation), and TDI (Time Domain Interpolation). The performance order of these estimation techniques is as follows: LPI, SCI, TDI, SOI, and LI. Also, LPI and SCI yield almost the same best performance in the low and middle SNR scenarios, while LPI outperforms SCI at the high SNR scenario. In terms of the complexity, TDI, LPI and SCI have roughly the same computational burden, while SOI and LI have less complexity. As a result, LPI and SCI are usually recommended because they yield the best trade-off between performance and complexity. When using different interpolation methods, the performance of various pilot patterns are different.

In this paper we consider a pilot pattern that is improved performance than the interpolation with LMMSE or LS that uses time-frequency correlation.

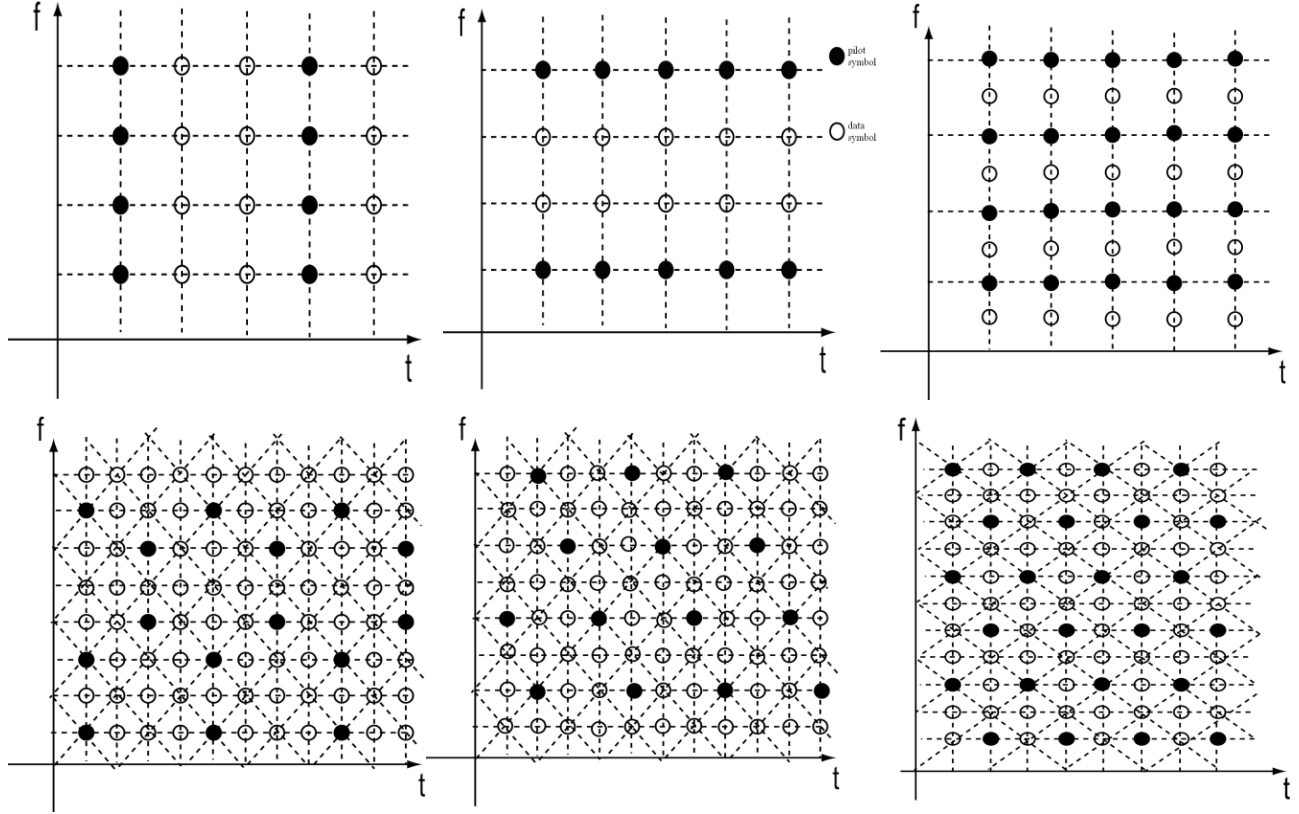

Figure 1. various pilot patterns (a) Block (b) Comb (c) Rectangular (d) Hexagonal (5) Parallelogram (6) Diamond

B. *Criteria for pilot pattern evaluation*

Usually evaluation criteria for pilot pattern are $d_t$ and $d_f$ for a given pilot density. This is appropriate when using the time- frequency correlation. When we consider non-causal interpolation, it is the distance between current estimation point and pilot symbol to affect the interpolation error. The distance between estimation point and the nearest pilot symbols is probably to be the largest factor which determines the interpolation error on that point. Based on this idea, definitions of maximum "distance" and average "distance" are defined as follows.

[Definition1]. Let us suppose that each coordinate axis of time-frequency plane is regularized. That is, suppose that $\overline{w_1}^{(4)} = \overline{w_2}^{(4)} = 1$. In this case, suppose that for a arbitrary data symbol point $D_s$ in a given pilot pattern, suppose that $d(D_s, p_i)$ is the "distance" between data symbol $D_s$ and pilot symbol $p_i$. Then $M = \arg\max_{D_s}[\min_i d(D_s, p_i)]$ is the point that gives the pilot pattern the maximum "distance" where $\min_i d(D_s, p_i)$ is defined as "maximum distance" $D_M$ of the pilot pattern. $E_S[\min_i d(D_s, p_i)]$ is defined as an "average distance" $D_E$ of the pilot pattern, where $E_S[X]$ is the expectation of $X$.

For a given pilot density, in order to make "maximum distance" $D_M$ to be the minimum, obviously, each pilot symbol should be uniformly spaced. In addition, for a given pilot density, if there is a pilot pattern in which maximum and average distance is minimal, interpolation errors might be smallest, and correlation based channel estimation may be improved for high correlation. In other words, for the equal error rate, we can decrease pilot density and, as a result, data rate will be increased.

## C. Optimal pilot pattern and its performance

### 1) Optimal pilot pattern

If we rotate the Diamond pilot pattern, a new pattern is obtained as follows.

In order to get uniformly spaced projection points of pilot symbols, $\theta$ and $\varphi$ of this pattern must satisfy following conditions:

$$\begin{cases} \cos\varphi = 3\cos(\theta+\varphi) \\ 3\sin\varphi = \sin(\theta+\varphi) \end{cases} \quad (3)$$

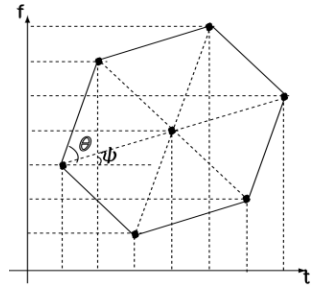

Figure 2. Pilot pattern which gives the optimal projection spacing both in time and in frequency axis.

Table1. Characteristics compared with several pilot pattern models. (Pilot density $D_P = \dfrac{4}{3\sqrt{3}} a^{-2}$)

|  | $D_M$ | $D_E$ | $d_t$ | $d_f$ |
|---|---|---|---|---|
| Comb | $0.658\,a^2$ | $0.325\,a^2$ | $1.299\,a^2$ | $1.299\,a^2$ |
| Block | $0.658\,a^2$ | $0.325\,a^2$ | $1.299\,a^2$ | $1.299\,a^2$ |
| Rotated rectangular | $0.806\,a$ | $0.437\,a$ | $1.194\,a$ | $1.194\,a$ |
| Rotated hexagonal | $a$ | $0.46\,a$ | $a/0.866\,a$ | $0.866\,a/a$ |
| Rotated diamond | $0.707\,a$ | $0.43\,a$ | $0.694\,a/0.401\,a$ | $0.401\,a/0.694\,a$ |
| Cell | $0.712\,a$ | $0.404\,a$ | $0.403\,a$ | $0.403\,a$ |

where $\theta = \arccos\dfrac{3}{5}, \varphi = \arctan\dfrac{1}{3}$

We will call this model "Cell pilot pattern" which is hexagonal consisting of six isosceles triangles with vertical angle of $\theta$ rotated by $\varphi$. $D_M, D_E, d_t, d_f$ of suggested model and other models are as follows:

The length of an edge in Rectangular, hexagonal or Diamond, and the length of equal sides in Cell, and the pilot spacing in time and frequency in Comb or Block is $a$. $a/0.866\,a$, $0.866\,a/a$ means that $d_t$ and $d_f$ are $a$ and

$0.866\,a$, respectively in case that projection points of pilot symbols are equally spaced in time axis and are $0.866\,a$ and $a$, respectively in case that projection points of pilot symbols are equally spaced in frequency axis. As can be seen in the table, in all aspects (except for $D_M$ of $0.707\,a$ in Diamond) Cell pattern seems to be superior to other patterns. If pilot density $D_p$ is constant, this pattern is superior to other patterns in terms of $d_t$ and $d_f$, affecting to accuracy of CIR (Channel Impulse Response) calculation as well as interpolation error, and in terms of $D_M$ and $D_E$, affecting to CIR calculation of the data symbol point obtained through interpolation. In general, for a given pilot density rotation on pilot pattern does not change the interpolation error. However, if the time-frequency correlation is used in CIR calculation on pilot symbol points or data points, such a rotation affects the whole error characteristics positively for its better characteristics of $d_t$ and $d_f$.

*2) Absorption area of pilot symbol in Cell pilot pattern and selection of pilot symbol set for interpolation.*

So far Cell pilot pattern was considered, however, these are actually only the parallelogram pilot pattern in view of time-frequency plane. In spite of that, these are considered as Cell type pattern because the set of pilot symbols that uses in CIR calculation at pilot symbol points and in CIR interpolation at data symbols depends on Cell pattern.

As in [12], a lot of works considered the interpolation

$$\hat{H}[n,k] = \sum_{p=-\infty}^{\infty}\sum_{q=-\infty}^{\infty} \tilde{H}_s[n+p, k+q] \cdot w[p,q] \qquad (4)$$

with non-causal filter, where $\hat{H}[n,k]$ is equal to $\tilde{H}[n,k]$ on the pilot symbol or zero otherwise, and $w[p,q]$ denotes the coefficient of the interpolator. However, these are not practical due to infinite computation quantity.

On the other hand, LI (linear interpolation) is difficult to use because the interpolation error is too large.

$$\hat{H}(ks+t) = \hat{H}_{LS}^P(k) + (\hat{H}_{LS}^P(k+1) - \hat{H}_{LS}^P(k)) \cdot t/s, \qquad (5)$$
$$0 \le t < s$$

where $s = N/N_p$ is sub-carrier spacing and $\hat{H}_{LS}^P(k)$ is LS estimated channel response at the pilot symbol.

Thus, we considered Cell pilot pattern which interpolates CIR at all the data symbols by using CIRs on 7 pilot symbols.

As can be seen in Fig.3, on every data symbol, 3 Cell patterns of pilot symbols which contain the point exist.

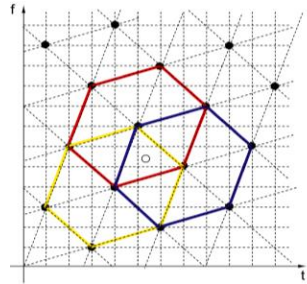

Figure 3.  Cell patterns of pilot symbol points which contain a data symbol point.

In this case, which pattern is to use is a problem. For this the absorption area of pilot symbol points is defined as following.

[Definition 2]

The set of data symbols which are closest to a pilot symbol in the regularized time-frequency plane is defined as absorption area of the pilot symbol.

The absorption area on every pilot pattern is as follows. In case that the absorption area is defined according to the definition 2, the pilot pattern for interpolation is chosen as following.

For every data symbol, a pilot symbol is determined to contain the data symbol in its absorption area, and selects a Cell pilot pattern centered on the pilot symbol. It is valid because the closer the data symbol is to the center of pilot pattern, the smaller interpolation error is. But if the data symbol is on the border of absorption area, interpolation error will be maximal. In this case, a pilot pattern, in which the data symbol point is on the right side as much as possible, must be selected to be the CIR calculation time to be minimal. When using pilot pattern adaptive method based on FASA that will be discussed, pilot pattern of which pilot density is large must be selected.

3) *CIR estimation in Cell pilot pattern*

When CIR on pilot symbol is calculated by LS method and LMMSE method, CIR on data symbol is calculated by following "distance" filter.

When the function values on N points $f(a_i), i = \overline{1, N}$ in 2-dimensional plane (time-frequency plane) are given, the function value on a point b is interpolated by

$$\hat{f}(b) = \frac{\sum_{i=1}^{N} \frac{1}{d_i(b)} f(a_i)}{\sum_{i=1}^{N} \frac{1}{d_i(b)}} \qquad (6)$$

where $d_i(b)$ is the "distance" between the point $a_i, i = 1 \sim N$ and the point b. Actually, $N$ is the number of pilot symbols which is used in interpolation and is 7, and the point b is the data symbol point inside Cell pilot pattern which is defined by 7 points. Essentially, the "distance" filter which is defined by the (6) is only the extension of the linear interpolation or bilinear interpolation. The "distance" interpolation by Rectangular pattern corresponds to bilinear interpolation and the "distance" interpolation on 1-dimension corresponds to linear interpolation. Equation (6) corresponds with the basic characteristics of linear interpolation that says the function value in known point is same as the original value and when the points are same distant from the known points, the function value is the arithmetic average of the known function values.

In fact,

$$\hat{f}(a_i) = \lim_{d_i(b) \to 0} \hat{f}(b) = \lim_{d_i(b) \to 0} \frac{\sum_{j=1}^{N} \frac{1}{d_j(b)} f(a_j)}{\sum_{j=1}^{N} \frac{1}{d_j(b)}} = \lim_{d_i(b) \to 0} \frac{\hat{f}(a_i) + \sum_{j=1}^{N} \frac{d_i(b)}{d_j(b)} f(a_j)}{1 + \sum_{j=1}^{N} \frac{d_i(b)}{d_j(b)}} = f(a_i) \qquad (7)$$

When the points are same distant, $d_i(b) = C$, then

$$f(m) = \frac{\sum_{j=1}^{N} \frac{1}{C} f(a_i)}{N \cdot \frac{1}{C}} = \frac{1}{N} \sum_{j=1}^{N} f(a_i) \qquad (8)$$

where m is the number of points which is same distant from the known points for interpolation.

The interpolation curved surface from the "distance" filter has bad peak-shaped characteristics in the center point area. If we call the curved surface area as A and other area as B, then in area A, the error from CIR estimation is large and the interpolation error is small and in area B vice versa.

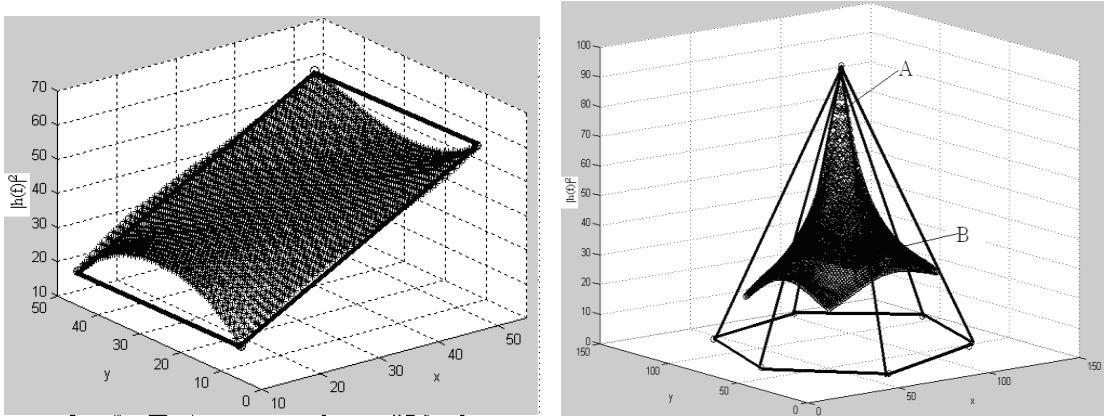

Figure 4. Interpolated curved surface using "distance" interpolation pattern     (a) Rectangular pilot pattern (b) Cell pilot pattern

As a result, the "distance" filter has the good performance in view of MSE criteria due to the relatively constant characteristics of the error in area A and B.

Equation (6) can be rewritten:

$$\hat{H}_d(t,f) = \frac{\sum d_i^{-1} H_i(t,f)}{\sum_i d_i^{-1}} \quad (9)$$

where $H_i(t,f)$ $i = \overline{1,7}$ : CIRs of the corresponding pilot symbol set.

$\hat{H}_d(t,f)$: CIR of the data symbol on which interpolation is going to be done.

Equation (9) can be rewritten:

$$\hat{H}_d(t,f) = \sum_{i=1}^{7} k_i \hat{H}_i^{\,p}(t,f) \quad (10)$$

where $k_i = d_i^{-1} \sum_{i=1}^{7} d_j^{-1}$, $\hat{H}_i^{\,p}(t,f)$ is CIR estimation value in $i^{th}$ pilot symbol.

The Mean Square Error (MSE) due to the CIR estimation error on pilot symbol point is as following:

$$MSE_{pilot} = E\left\{\left[\sum k_i (\hat{H}_i^{(p)}(t,f) - H_i^{(p)}(t,f))\right]^2\right\}$$
$$= E\left\{(\sum k_i z_i)^2\right\} = E\left\{\sum k^2_i z^2_i\right\} = \sigma_n^2 \sum_i E\{k_i^2\} \quad (11)$$

Where $H_i^{\,p}(t,f)$: actual CIR in pilot symbol. $z_i$: CIR estimation error with $N(0,\sigma_n^2)$ that satisfies

$$\hat{H}_i^{(p)}(t,f) = H_i^{(p)}(t,f) + z_i$$

Using the probability density function of $k_i$, $\sum E[k_i^2]$ is about 0.15.

Therefore, with the "distance" filter, the pilot symbol estimation error is 8 dB. Of course, "distance" interpolation has larger computational burden than LI and SOI, but is only due to the number of pilot symbols in interpolation, and as can be seen in (9), it doesn't include any computation of vector, matrix and the second-order operation; thus, computational burden is smaller than other interpolation schemes like LPI

On performance, when considering LPI and SCI yield almost the same performance in low and middle SNR scenarios, "distance" filter is better than LPI. The "distance" filter is appropriate for a filter in the first-order interpolation with the pilot pattern due to the robustness against large errors of known function values $H_i(t,f)$ and less computational burden.

### III. THE RESULTS OF SIMULATION

The performance of Cell pilot pattern suggested and the "distance" filter using it is proved through simulation.

SNR-BER curve of parallelogram pilot pattern in Fig.5 is the result of using the same interpolation suggested in literature [19]. This interpolation method uses pattern and performs interpolation along the time or frequency axis with less change. As can be seen in Fig.5, Cell pilot pattern provides 10 dB than Comb (SNR is below 15 dB), 1dB than Rectangular, about 13dB BER benefits than parallelogram. The SNR-BER curve in Fig.5 is BER from which BER at $SNR = \infty$ is subtracted. Although it has no noise, BER (called $BER_\infty$) at $SNR = \infty$ can be estimated as an existing BER by pilot pattern itself. Now the plots in Fig.5 only show BER due to effect of noise t o the pilot pattern. Total BER ($BER_{Total}$) with SNR is plotted in Fig.6.

As can be seen in Fig.6, Cell type ("distance" interpolation) has lower BER than other pilot patterns using TDI and correlation method in the section of SNR of less than 10dB. The method in literature [13] is ML and Approximate LMMSE estimation on the pilot symbol, but it has very large computation burden due to matrix calculation.). We compare the bilinear and "distance" interpolation on the same pattern (Rectangular) as following.

As can be seen, "distance" interpolation filter has lower BER than other first order interpolation. ("Distance" interpolation has approximately 1dB BER benefit than bilinear interpolation.

Table2. Simulation conditions

| № | parameter | value |
|---|---|---|
| 1 | Number of sub carriers | 128 |
| 2 | Spacing of sub carriers $\Delta f$ | 125 kHz |
| 3 | FFT size | 128 |
| 4 | Modulation | QPSK,8QAM,16QAM |
| 5 | Cyclic Prefix | 16(Samples) |
| 6 | Channel | AWGN |
| 7 | Pilot Density | 5.7% |
| 8 | Frequency Response Normal Variance | 0.177,0.32,0.357 |
| 9 | Normalized Doppler frequency( kHz) | 0.02455 |
| 10 | SNR | 0~25dB |

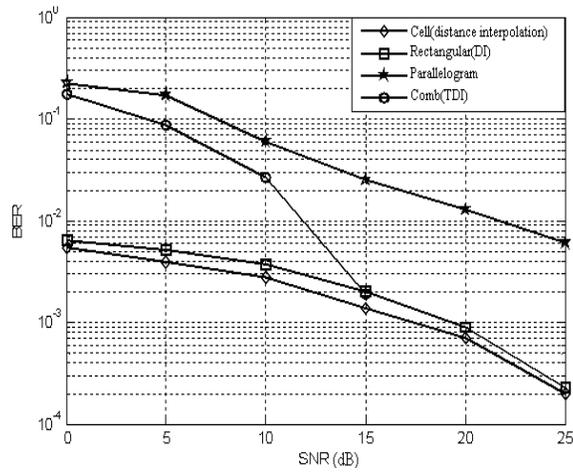

Figure 5. BER vs. SNR of the "distance" interpolation for each pilot pattern.

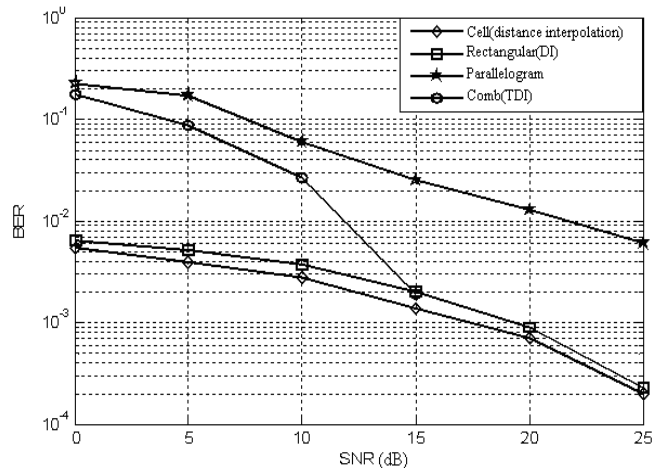

Figure 6. BER vs. SNR of the "distance" interpolation for each pilot pattern.

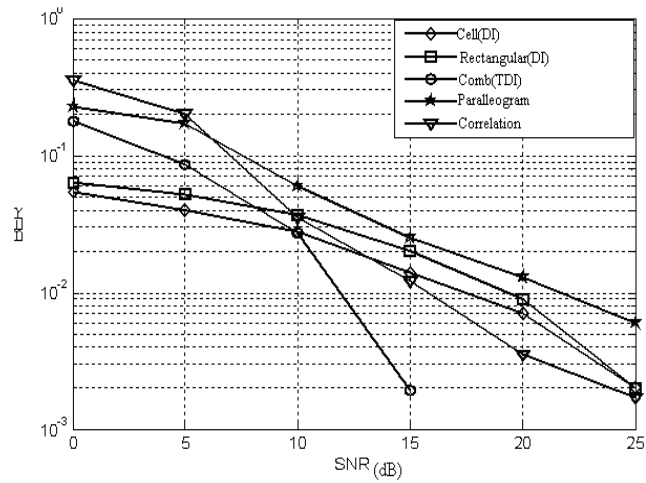

Figure 7. BERTOTAL vs. SNR in "distance" interpolation

IV. CONCLUSION

In this paper, we proposed optimal pilot pattern, new criteria of pilot pattern evaluation "maximum distance" and "average distance" and distance filter suitable to Cell pilot pattern in downlink OFDM communication systems. The simulation results show that for a given pilot density, BER of Cell pilot pattern is superior to other pilot patterns and the combination of Cell pilot pattern and "distance" filter provides better performance than the performance of LMMSE estimation which implements time-frequency correlation with some patterns.

BIOGRAPHIES

**Yong Chan Lee** received his master of computer engineering in 2013. His research interest is in the area of WCDMA and

OFDM communication system.

**Won Chol Jang** received his master of computer engineering in 2015. His research interest is in the area of OFDM communication system.

**Un Kyong Choe** received her master of computer engineering in 2012. Her research interest is in the area of OFDM communication system.

**Gyong Chol Leem** received her master of computer engineering in 2008. Her research interest is in the area of WCDMA communication system.